\documentclass[10pt,a4paper,twoside]{article}
\usepackage{epsfig}
\usepackage{baltlat6}
\usepackage{array}
\usepackage{here}
\begin{document}
\ \
\vspace{0.5mm}
\setcounter{page}{613}
\vspace{8mm}

\titlehead{Baltic Astronomy, vol.\,20, 613--617, 2011}

\titleb{STARK BROADENING OF In III LINES IN ASTROPHYSICAL
\\ AND LABORATORY PLASMA}

\begin{authorl}
\authorb{Z. Simi\' c}{1}
\authorb{M. S. Dimitrijevi\' c}{1}
\authorb{A. Kova\v cevi\' c}{2} and
\authorb{S. Sahal-Br\' echot}{3}
\end{authorl}

\begin{addressl}
\addressb{1}{Astronomical Observatory, Volgina 7, 11060 Belgrade,Serbia;
\\ zsimic@aob.bg.ac.rs; mdimitrijevic@aob.bg.ac.rs}
\addressb{2}{Department for Astronomy, Faculty for Mathematics,
Studentski Trg 16,
\\ 11000 Belgrade, Serbia; andjelka@matf.bg.ac.rs}
\addressb{3}{Observatoire de Paris, 92195 Meudon Cedex, France;
\\ sylvie.sahal-brechot@obspm.fr}
\end{addressl}

\submitb{Received: 2011 August 8; accepted: 2011 August 15}

\begin{summary} Besides the need of Stark broadening parameters for a number of problems in physics,
and plasma technology, in hot star atmospheres the conditions exist where Stark widths are comparable
and even larger than the thermal Doppler widths. Using the semiclassical perturbation method we investigated here
the influence of collisions with charged particles for In III spectral lines.
We determined a number of Stark broadening parameters
important for the investigation of plasmas in the atmospheres of A-type stars and white dwarfs.  Also, we have compared the
obtained results with existing experimental data. The results will be included in the STARK-B
database, the Virtual Atomic and Molecular Data Center and the Serbian Virtual Observatory.
\end{summary}


%% \resthead is the RUNNING TITLE at top of the pages
\resthead{Stark Broadening of In III lines}
{Z. Simi\' c, M. S. Dimitrijevi\'c, A. Kova\v cevi\'c, S. Sahal-Br\'echot}

\sectionb{1}{INTRODUCTION}
Stark broadening parameters of indium spectral lines are of interest for a number of
problems in astrophysics, physics, and plasma technology. For example indium is identified
in the spectrum of HD110066, an A-type chemically peculiar star
(Cowley et al. 1974).

Dimitrijevi\' c \& Sahal-Br\' echot (1999) determined, using semiclassical perturbation theory
(Sahal-Br\' echot 1969a,b) Stark broadening parameters due to collisions
with electrons, protons and helium ions of 20 In III multiplets, for the temperatures
from 20000 K to 50000 K. Djeni\v ze et al. (2006) obtained experimentally Stark widths for sixteen
In III spectral lines at a temperature of 13000 K without citation of
Dimitrijevi\' c \& Sahal-Br\' echot (1999).

In order to make a better comparison with their experimental results and to provide the data of
interest for hot stellar plasma research, we recalculated Stark broadening parameters for ten In III spectral lines using the semiclassical perturbation method
(Sahal-Br\' echot 1969a,b), within four previously considered multiplets
(Dimitrijevi\' c \& Sahal-Br\' echot 1999), for which the experimental data
(Djeni\v ze et al. 2006) exist. Differently with the previous work, we calculated the data for each line separately, and we increased  the temperature range from 20 000 K - 50 000 K to 10 000 K - 100 000 K to cover the temperature of the experimental investigation. The obtained results are used to investigate the influence of Stark broadening of the In III lines in the A-type
stellar atmospheres.

\sectionb{2}{RESULTS AND DISCUSSIONS}

The calculations have been performed within the semiclassical
perturbation formalism,  developed and discussed in detail in
Sahal-Br\' echot (1969a,b). This formalism, as well as the corresponding
computer code, have been optimized and updated several times
(see e.g. Sahal-Br\' echot 1974; Dimitrijevi\' c \& Sahal-Br\' echot 1984; Dimitrijevi\' c et al. 1991; and the review by Dimitrijevi\' c 1996).
The needed atomic energy levels for Stark broadening calculations
were taken from Bhatia (1978).

Table 1. shows the Stark broadening parameters for electron, proton, and helium impacts
(full widths at half intesity maximum, and shifts) for ten In III lines, obtained by using the semiclassical
perturbation method for a perturber density of 10$^{17}$cm$^{-3}$ and temperatures from 10 000 K up to 100 000 K.
The quantity C (given in \AA\ cm$^{-3}$), when divided by the corresponding FWHM,
gives an estimate for the maximum perturber density for which tabulated data may be used.

In Table 2. we show a comparison between experimental (Djeni\v ze et al. 2006) and our theoretical results for
Stark line widths. One can see a large disagreement between both results.
Ratio of experimental and theoretical widths changes from 0,25 to 0.52,
i.e. the experimental widths are two to four times smaller than theoretical. A new experimental determination
of Stark widths for In III spectral lines would be important.

In order to investigate the influence of Stark broadening mechanism for In III lines in stellar plasma conditions,
we calculated Stark widths for the Kurucz's (1979) model of the A-type
atmosphere with $T_{eff}$ = 10 000 K and log $g$ = 4.5
and compared them with the Doppler broadening (see Figure 1). We found that the photospheric layers exist where the
Doppler and Stark widths are comparable and even where the Stark width is dominant.

\begin{figure}[!h]
\vbox{
\centerline{\psfig{figure=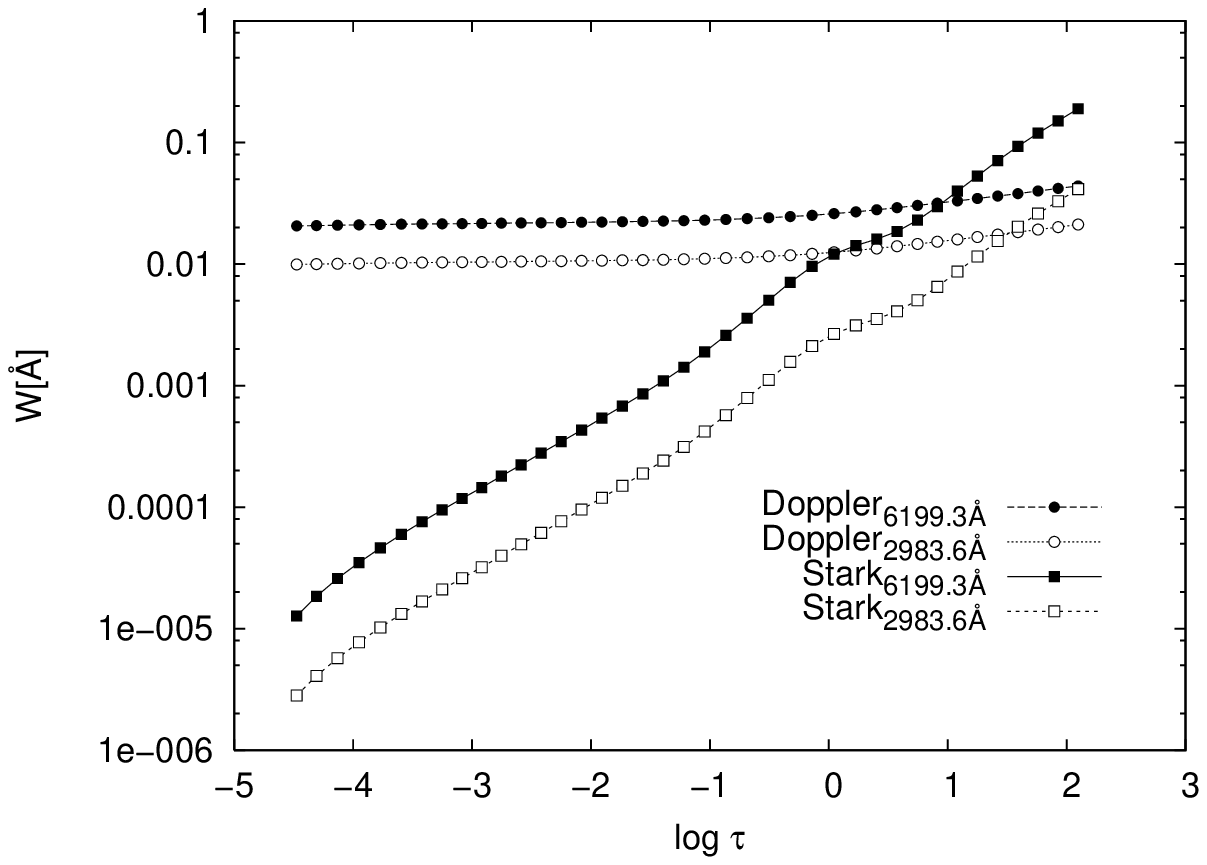,width=96mm,angle=0,clip=}}
\vspace{1mm}
\captionb{1}
{The Doppler and Stark widths for two In III lines at 2983.6 and 6199.3 \AA\
as functions of the Rosseland optical depth for an A-type star atmosphere ($T_{eff}$ = 10 000 K, log $g$ = 4.5).
One can see that the Stark broadening mechanism is absolutely dominant in comparison
with the thermal Doppler mechanism in deeper layers of the stellar atmosphere (log ${\tau} > 1.0$). }
}
\end{figure}

\begin{figure}[!h]
\vbox{
\centerline{\psfig{figure=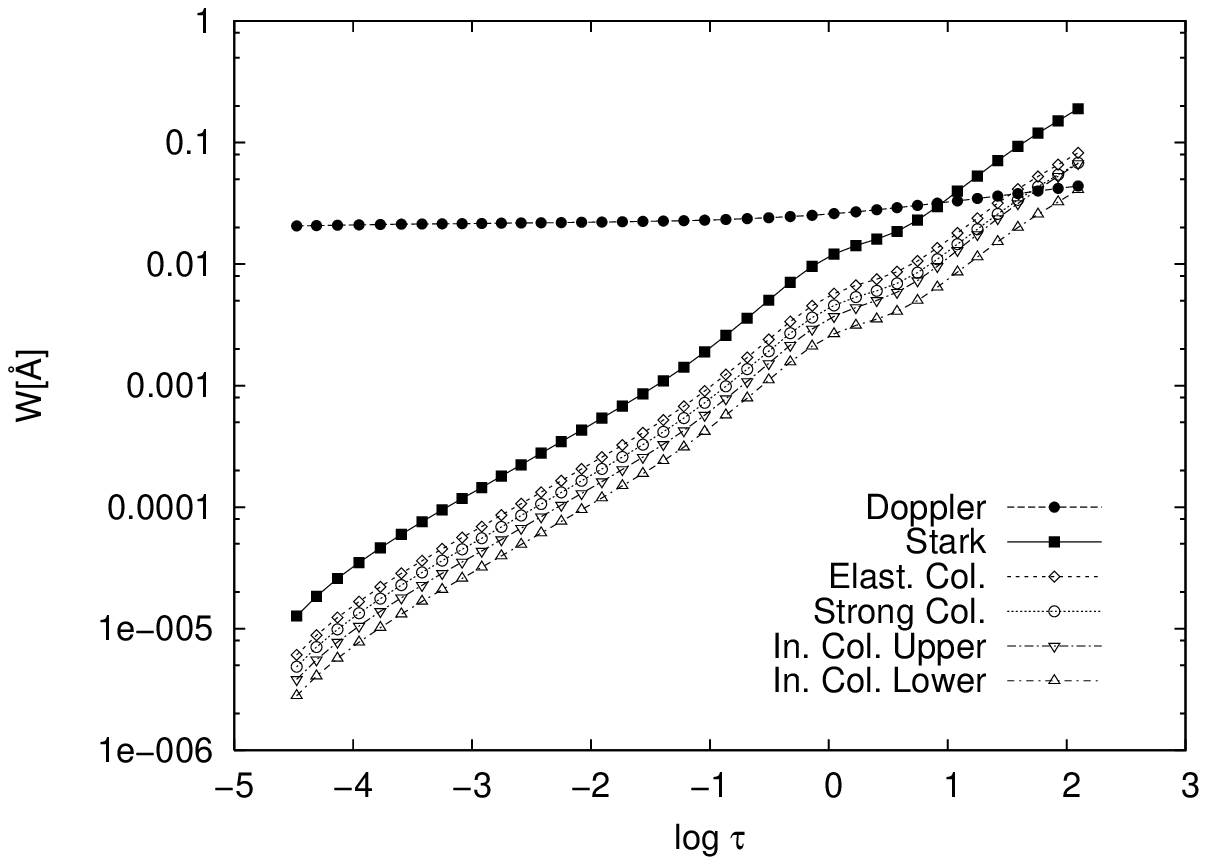,width=96mm,angle=0,clip=}}
\vspace{1mm}
\captionb{2}
{The thermal (Doppler), total Stark width and contributions of different collisional processes to the total Stark width for the
In III 6199.3 \AA\ line as a function of the Rosseland optical depth for a model atmosphere of A-type star
($T_{eff}$ = 10000 K, log $g$ = 4.5).}
}
\end{figure}

Additionally, we compared thermal Doppler and the total Stark widths and the contribution of different
collisional processes for
In III 6199.3 \AA\ line for the same model atmosphere.
The contribution of elastic collisions is leading in the total Stark widths (see Figure 2).
The strong collisions, inelastic collision inducing transition from or to upper atomic energy levels and inelasic
collisions for lower levels are less important.

The obtained Stark broadening parameters will supplement the STARK-B database
(http://stark-b.obspm.fr),
dedicated for modelling of stellar atmospheres, analysis and synthesis of stellar spectra,
as well as for investigations of laboratory plasma, inertial fusion plasma,
laser development and for plasmas in technology.
This database is a part of a FP7 project Virtual Atomic and Molecular Data Center - VAMDC
(P.I. Marie Lise Dubernet), with the following aims: (i) to build a secure,
flexible and interoperable e-science environment based interface to the existing Atomic and Molecular
databases; (ii) to coordinate groups involved in the generation, evaluation, and use of atomic and molecular
data, and (iii) to  provide a forum for training of potential users (Dubernet et al. 2010, Rixon et al. 2011).

\thanks{ This work has been supported by VAMDC,  funded under the "Combination of Collaborative Projects and
Coordination and  Support Actions" Funding Scheme of The Seventh Framework Program. Call topic:
INFRA-2008-1.2.2 Scientific Data Infrastructure. Grant Agreement number: \break
239108.  The authors are also  grateful for the support  provided by Ministry of Education and Science of
Republic of Serbia through projects  176002 "Influence of collisional processes on astrophysical plasma spectra",
and III44002 "Astroinformatics: Application of IT in astronomy and related disciplines".}

\begin{table}[]
\begin{center}
\vbox{\footnotesize\tabcolsep=3pt
\parbox[c]{124mm}{\baselineskip=10pt
{\smallbf\ \ Table 1.}{\small\
Stark broadening parameters: widths (FWHM) and shifts for In III spectral lines obtained
within semiclasical approach for a perturber density of 10$^{17}$ cm$^{-3}$ and temperatures
from 10000 K up to 100000 K.\lstrut}}
\vspace*{0.5cm} \footnotesize \scriptsize
\begin{tabular}{c c c c c c c c}
\hline \hline
  & & & & & & & \\
Transition & {\rm T(K)} & {\rm W$_{\mathrm {e^{-}}}$(\AA)} & {\rm d$_{\mathrm {e^{-}}}$(\AA)} & {\rm W$_{\mathrm {p^{+}}}$(\AA)}
& {\rm d$_{\mathrm {p^{+}}}$(\AA)} & {\rm W$_{\mathrm {He^{+}}}$(\AA)} & {\rm d$_{\mathrm {He^{+}}}$(\AA)} \\
  & & & & & & & \\
\hline
                            &  10000 & 0.358 & 0.637E-02 & 0.118E-01 & 0.244E-02 & 0.158E-01 & 0.232E-02 \\
{\rm In III}                &  13000 & 0.319 & 0.698E-02 & 0.144E-01 & 0.317E-02 & 0.181E-01 & 0.291E-02 \\
${\mathrm {5d\ ^{2}D_{3/2}-4f\ ^{2}F_{5/2}^{o}}}$ &  20000 & 0.268 & 0.632E-02 & 0.181E-01 & 0.451E-02 & 0.222E-01 & 0.401E-02 \\
2983.6 \AA                  &  30000 & 0.231 & 0.608E-02 & 0.220E-01 & 0.575E-02 & 0.242E-01 & 0.500E-02 \\
{\rm C= 0.77E+20}           &  50000 & 0.197 & 0.680E-02 & 0.248E-01 & 0.747E-02 & 0.267E-01 & 0.626E-02 \\
                            & 100000 & 0.167 & 0.636E-02 & 0.282E-01 & 0.921E-02 & 0.297E-01 & 0.748E-02 \\
\hline
                            &  10000 & 0.365 & 0.638E-02 & 0.120E-01 & 0.244E-02 & 0.161E-01 & 0.232E-02 \\
{\rm In III}                &  13000 & 0.326 & 0.695E-02 & 0.147E-01 & 0.316E-02 & 0.185E-01 & 0.291E-02 \\
${\mathrm {5d\ ^{2}D_{5/2}-4f\ ^{2}F_{5/2}^{o}}}$ &  20000 & 0.273 & 0.626E-02 & 0.185E-01 & 0.450E-02 & 0.226E-01 & 0.401E-02 \\
3009.7 \AA                  &  30000 & 0.236 & 0.602E-02 & 0.225E-01 & 0.575E-02 & 0.247E-01 & 0.500E-02 \\
{\rm C= 0.79E+20}           &  50000 & 0.201 & 0.672E-02 & 0.252E-01 & 0.747E-02 & 0.272E-01 & 0.627E-02 \\
                            & 100000 & 0.170 & 0.627E-02 & 0.287E-01 & 0.921E-02 & 0.303E-01 & 0.751E-02 \\
\hline
                            &  10000 & 0.365 & 0.639E-02 & 0.120E-01 & 0.244E-02 & 0.161E-01 & 0.233E-02 \\
{\rm In III}                &  13000 & 0.326 & 0.696E-02 & 0.147E-01 & 0.317E-02 & 0.184E-01 & 0.291E-02 \\
${\mathrm {5d\ ^{2}D_{5/2}-4f\ ^{2}F_{7/2}^{o}}}$ &  20000 & 0.273 & 0.628E-02 & 0.185E-01 & 0.451E-02 & 0.226E-01 & 0.402E-02 \\
3008.9 \AA                  &  30000 & 0.236 & 0.604E-02 & 0.225E-01 & 0.576E-02 & 0.247E-01 & 0.501E-02 \\
{\rm C= 0.78E+20}           &  50000 & 0.201 & 0.674E-02 & 0.252E-01 & 0.749E-02 & 0.272E-01 & 0.628E-02 \\
                            & 100000 & 0.170 & 0.629E-02 & 0.287E-01 & 0.923E-02 & 0.303E-01 & 0.753E-02 \\
\hline
                            &  10000 &  1.26 & -0.112     & 0.426E-01 & -0.133E-01 & 0.554E-01 & -0.122E-01 \\
{\rm In III}                &  13000 &  1.12 & -0.701E-01 & 0.516E-01 & -0.167E-01 & 0.640E-01 & -0.153E-01 \\
${\mathrm {6s\ ^{2}S_{1/2}-6p\ ^{2}P_{1/2}^{o}}}$ &  20000 & 0.946 & -0.710E-01 & 0.649E-01 & -0.231E-01 & 0.774E-01 & -0.196E-01 \\
5250.3 \AA                  &  30000 & 0.824 & -0.599E-01 & 0.782E-01 & -0.287E-01 & 0.841E-01 & -0.248E-01 \\
{\rm C= 0.48E+21}           &  50000 & 0.721 & -0.635E-01 & 0.876E-01 & -0.363E-01 & 0.929E-01 & -0.294E-01 \\
                            & 100000 & 0.626 & -0.622E-01 & 0.100     & -0.436E-01 & 0.103     & -0.355E-01 \\
\hline
                            &  10000 &  1.49 & -0.158     & 0.472E-01 & -0.177E-01 & 0.616E-01 & -0.161E-01 \\
{\rm In III}                &  13000 &  1.33 & -0.135     & 0.575E-01 & -0.221E-01 & 0.711E-01 & -0.201E-01 \\
${\mathrm {6s\ ^{2}S_{1/2}-6p\ ^{2}P_{3/2}^{o}}}$ &  20000 &  1.12 & -0.118     & 0.727E-01 & -0.303E-01 & 0.867E-01 & -0.258E-01 \\
5646.9 \AA                  &  30000 & 0.967 & -0.867E-01 & 0.883E-01 & -0.378E-01 & 0.944E-01 & -0.324E-01 \\
{\rm C= 0.51E+21}           &  50000 & 0.842 & -0.888E-01 & 0.996E-01 & -0.470E-01 & 0.105     & -0.381E-01 \\
                            & 100000 & 0.726 & -0.852E-01 & 0.115     & -0.563E-01 & 0.117     & -0.459E-01 \\
\hline
                            &  10000 & 1.09  & 0.461 & 0.386E-01 & 0.389E-01 & 0.418E-01 & 0.324E-01 \\
{\rm In III}                &  13000 & 0.996 & 0.408 & 0.454E-01 & 0.455E-01 & 0.502E-01 & 0.387E-01 \\
${\mathrm {6p\ ^{2}P_{1/2}^{o}-7s\ ^{2}S_{1/2}}}$ &  20000 & 0.860 & 0.338 & 0.630E-01 & 0.591E-01 & 0.625E-01 & 0.481E-01 \\
4024.8 \AA                  &  30000 & 0.785 & 0.285 & 0.763E-01 & 0.689E-01 & 0.711E-01 & 0.564E-01 \\
{\rm C= 0.13E+21}           &  50000 & 0.728 & 0.251 & 0.918E-01 & 0.816E-01 & 0.825E-01 & 0.667E-01 \\
                            & 100000 & 0.666 & 0.196 & 0.114     & 0.976E-01 & 0.982E-01 & 0.777E-01 \\
\hline
                            &  10000 &  1.19 & 0.491 & 0.435E-01 & 0.426E-01 & 0.474E-01 & 0.354E-01 \\
{\rm In III}                &  13000 &  1.09 & 0.434 & 0.511E-01 & 0.498E-01 & 0.568E-01 & 0.424E-01 \\
${\mathrm {6p\ ^{2}P_{3/2}^{o}-7s\ ^{2}S_{1/2}}}$ &  20000 & 0.942 & 0.358 & 0.705E-01 & 0.648E-01 & 0.703E-01 & 0.528E-01 \\
4253.8 \AA                  &  30000 & 0.864 & 0.302 & 0.853E-01 & 0.757E-01 & 0.799E-01 & 0.619E-01 \\
{\rm C= 0.15E+21}           &  50000 & 0.804 & 0.267 & 0.102     & 0.896E-01 & 0.920E-01 & 0.729E-01 \\
                            & 100000 & 0.741 & 0.208 & 0.126     & 0.108     & 0.110     & 0.852E-01 \\
\hline
                              &  10000 &  1.62 & -0.358E-01 & 0.681E-01 & -0.112E-01 & 0.876E-01 & -0.106E-01 \\
{\rm InIII}                   &  13000 &  1.45 & -0.365E-01 & 0.811E-01 & -0.145E-01 & 0.101     & -0.133E-01 \\
${\mathrm {5d\ ^{2}D_{3/2}-6p\ ^{2}P_{1/2}^{o}}}$ &  20000 &  1.23 & -0.426E-01 & 0.102     & -0.206E-01 & 0.120     & -0.182E-01 \\
6199.3 \AA                    &  30000 &  1.06 & -0.372E-01 & 0.120     & -0.261E-01 & 0.130     & -0.228E-01 \\
{\rm C= 0.61E+21}             &  50000 & 0.916 & -0.456E-01 & 0.133     & -0.339E-01 & 0.142     & -0.283E-01 \\
                              & 100000 & 0.781 & -0.449E-01 & 0.149     & -0.416E-01 & 0.157     & -0.339E-01 \\
\hline
                              &  10000 &  1.35 & -0.173E-01 & 0.604E-01 & -0.685E-02 & 0.774E-01 &-0.658E-02 \\
{\rm InIII}                   &  13000 &  1.21 & -0.169E-01 & 0.716E-01 & -0.896E-02 & 0.895E-01 &-0.836E-02 \\
${\mathrm {5d\ ^{2}D_{3/2}-6p\ ^{2}P_{3/2}^{o}}}$ &  20000 &  1.03 & -0.233E-01 & 0.895E-01 & -0.129E-01 & 0.105     &-0.117E-01 \\
5724.6 \AA                    &  30000 & 0.896 & -0.210E-01 & 0.105     & -0.170E-01 & 0.114     &-0.145E-01 \\
{\rm C= 0.56E+21}             &  50000 & 0.776 & -0.271E-01 & 0.116     & -0.219E-01 & 0.125     &-0.187E-01 \\
                              & 100000 & 0.666 & -0.269E-01 & 0.130     & -0.275E-01 & 0.137     &-0.225E-01 \\
\hline
                              &  10000 &  1.40 & -0.187E-01 & 0.626E-01 & -0.728E-02 & 0.801E-01& -0.698E-02 \\
{\rm InIII}                   &  13000 &  1.26 & -0.182E-01 & 0.741E-01 & -0.951E-02 & 0.927E-01& -0.885E-02 \\
${\mathrm {5d\ ^{2}D_{5/2}-6p\ ^{2}P_{3/2}^{o}}}$ &  20000 &  1.07 & -0.248E-01 & 0.928E-01 & -0.136E-01 & 0.109    & -0.124E-01 \\
5821.2 \AA                    &  30000 & 0.929 & -0.224E-01 & 0.109     & -0.180E-01 & 0.118    & -0.154E-01 \\
{\rm C= 0.57E+21}             &  50000 & 0.805 & -0.288E-01 & 0.120     & -0.231E-01 & 0.129    & -0.198E-01 \\
                              & 100000 & 0.692 & -0.287E-01 & 0.135     & -0.290E-01 & 0.142    & -0.237E-01 \\
\hline \hline \normalsize
\end{tabular}
}
\end{center}
\vskip-8mm
\end{table}

\begin{table}[]
\begin{center}
\vbox{\footnotesize\tabcolsep=3pt
\parbox[c]{124mm}{\baselineskip=10pt
{\smallbf\ \ Table 2.}{\small\
Comparison between experimental-W$_{\mathrm {m}}$ and theoretical-W$_{\rm th}$ results.\lstrut}}
\vspace*{0.5cm} \footnotesize \scriptsize
\begin{tabular}{cccc}
\hline \hline
Transition &
$\lambda_{\mathrm {m}}$ (\AA) &
{\rm W$_{\mathrm {m}}$(\AA)} &
${\mathrm {{W_{m}}\over{W_{th}}}}$ \\
\hline
${\mathrm {5d\ ^{2}D_{3/2}-4f\ ^{2}F_{5/2}^{o}}}$  & 2982.8 & 0.240 & 0.51  \\
\hline
${\mathrm {5d\ ^{2}D_{5/2}-4f\ ^{2}F_{5/2}^{o}}}$  & 3008.1 & 0.250 & 0.52  \\
\hline
${\mathrm {5d\ ^{2}D_{5/2}-4f\ ^{2}F_{7/2}^{o}}}$  & 3008.8 & 0.210 & 0.44  \\
\hline
${\mathrm {6s\ ^{2}S_{1/2}-6p\ ^{2}P_{1/2}^{o}}}$  & 5248.8 & 0.717 & 0.43  \\
\hline
${\mathrm {6s\ ^{2}S_{1/2}-6p\ ^{2}P_{3/2}^{o}}}$  & 5645.2 & 0.520 & 0.27  \\
\hline
${\mathrm {6p\ ^{2}P_{1/2}^{o}-7s\ ^{2}S_{1/2}}}$  & 4023.8 & 0.545 & 0.37  \\
\hline
${\mathrm {6p\ ^{2}P_{3/2}^{o}-7s\ ^{2}S_{1/2}}}$  & 4252.7 & 0.480 & 0.30  \\
\hline
${\mathrm {5d\ ^{2}D_{3/2}-6p\ ^{2}P_{1/2}^{o}}}$  & 6197.7 & 0.800 & 0.37  \\
\hline
${\mathrm {5d\ ^{2}D_{3/2}-6p\ ^{2}P_{3/2}^{o}}}$  & 5723.2 & 0.450 & 0.25  \\
\hline
${\mathrm {5d\ ^{2}D_{5/2}-6p\ ^{2}P_{3/2}^{o}}}$  & 5819.5 & 0.520 & 0.28  \\
\hline \hline \normalsize
\end{tabular}
}
\end{center}
\vskip-8mm
\end{table}

\References

\refb Bhatia, K.S., 1978, J. Phys. B, Vol. 11, No 14, 2421

\refb Cowley, C.R., Hartoog, M.R., \& Cowley, A.P., 1974, ApJ, 194, 343

\refb Dimitrijevi\'c, M.S., 1996, Zh. Priklad. Spektrosk., 63, 810

\refb Dimitrijevi{\' c}, M.S., \& Sahal-Br{\' e}chot, S., 1984, JQSRT, 31, 301

\refb Dimitrijevi{\' c}, M.S., Sahal-Br{\' e}chot, S., \& Bomier, V., 1991, A\&AS, 89, 581

\refb Dimitrijevi{\' c}, M.S., \& Sahal-Br{\' e}chot, S., 1999, Journal of Applied Spectroscopy, Vol 66, No. 6, 868

\refb Djeni\v ze, S., Bukvi\'c, S., Sre\'ckovi\'c, A., \& Nikoli\'c, Z., 2006, Spectrochimica Acta, Part B, 61, 588

\refb Dubernet M. L., Boudon V., Culhane J. L. et al., 2010, JQSRT  111, 2151

\refb Kurucz, R.L., 1979, ApJS, 40, 1

\refb  Rixon G., Dubernet M. L., Piskunov N. et al., 2011, AIP Conference Proceedings 1344, 107

\refb Sahal-Br{\' e}chot, S., 1969a, A\&A, 1, 91

\refb Sahal-Br{\' e}chot, S., 1969b, A\&A, 2, 322

\refb Sahal-Br{\' e}chot, S., 1974, A\&A, 35, 321

\end{document}